\documentclass[fleqn,usenatbib]{mnrsas}

\usepackage{newtxtext,newtxmath}

\usepackage[T1]{fontenc}

\DeclareRobustCommand{\VAN}[3]{#2}
\let\VANthebibliography\thebibliography
\def\thebibliography{\DeclareRobustCommand{\VAN}[3]{##3}\VANthebibliography}

\usepackage{graphicx}	
\usepackage{amsmath}	

\title[Detection of extraterr. civilisation]{Detection of an extraterrestrial technical civilisation on the extrasolar planet GJ\,1132\,b
}

\author[Hessman, Cameron \& Horne]{
Frederic V. Hessman,$^{1}$\thanks{E-mail: fhessma@uni-goettingen.de}
Andrew {Collier\,Cameron},$^{2}$
and Keith Horne$^{2}$
\\
$^{1}$Institut~f\"ur~Astrophysik und Geophysik, University~of~G\"ottingen, Friedrich-Hund-Platz 1, G\"ottingen, Germany \\
$^{2}$SUPA School of Physics \& Astronomy, St~Andrews University, St~Andrews KY16~9NS, Scotland UK\\
}

\date{Not acceptable; received April 1}

\pubyear{2025}

\begin{document}
\label{firstpage}
\pagerange{\pageref{firstpage}--\pageref{lastpage}}
\maketitle

\begin{abstract}
We report the detection of whisky in the atmosphere of the extrasolar super-Earth planet GJ\,1132\,b from transmission spectroscopic data.
It is seen both in atmospheric absorption as well as in chromospheric emission, the latter probably due to the intense heating of the co-rotating planet's day-side surface.
This detection cannot be explained using natural sources of alcohol, implying that there must be a technically advanced civilisation -- possibly originating from the neighboring habitable planet GJ\,1132\,c -- that is engaged in massive distilling operations accompanied by high levels of industrial pollution.
The reason for the necessarily vast scale of production is either to produce rocket fuel for an interplanetary economy or, more likely, for an unusually high level of personal consumption.
The latter hypothesis suggests a novel explanation for the Fermi Paradox (the lack of indirect or direct contact with extraterrestrials): a technically versed civilisation would be incapable of achieving the higher technical levels necessary for the development of a detectable radio signature -- much less interstellar travel -- at the suggested rates of consumption.
\end{abstract}




\section{Introduction}

Transmission spectroscopy, a pivotal tool in astrophysics, elucidates the atmospheric compositions of exoplanets by analyzing the light passing through or around these distant worlds during transits. \cite{Charbonneau2002} initiated this field with the groundbreaking observation of the first exoplanetary transit, opening the door to subsequent studies such as \cite{Kreidberg2014}, who explored the atmosphere of GJ\,1214\,b using the Hubble Space Telescope. The imminent launch of the James Webb Space Telescope \citep{Beichman2014} promises further advancements, highlighting the continuous evolution of this technique. 

In addition to deciphering atmospheric compositions, transmission spectroscopy plays a crucial role in the search for signs of life beyond our solar system. \cite{Seager2005} investigated the detectability of biomarkers like oxygen and ozone, presenting the potential for identifying habitable environments. \cite{Kaltenegger2009} extended this concept to habitable zones, emphasizing the significance of transmission spectroscopy in characterizing life-friendly conditions. As technology advances, the possibility of detecting biosignatures becomes increasingly feasible, marking transmission spectroscopy as a key player in the exploration of extraterrestrial life.

Beyond the search for life, transmission spectroscopy has ventured into the realm of technosignatures—indicators of advanced technological civilizations. \cite{Schindler2018} proposed using transmission spectroscopy to identify artificial pollutants in exoplanet atmospheres, while \cite{Wright2014} discussed the potential of large telescopes to survey for technosignatures. This pioneering approach adds a fascinating dimension to the quest for intelligent extraterrestrial life, showcasing the versatility and expanding applications of transmission spectroscopy in the ever-evolving field of exoplanetary science.

As part of the Scottish-German research project ``Buaidh Astrofiosaig Uisge-Beatha Braiche'', we have considered whether distilled alcohol might be detected in the atmospheres of extrasolar planets.

\section{The absorption coefficient of distilled alcohols}
\label{sec:abs}

In the commercial literature, there have been many attempts at characterizing the wavelength-dependent absorbance of distilled alcohol (hereafter abbreviated as ``A''), primarily for the purpose of quality control and authentication \citep{wojcicki2015, okolo2023}.
The mass-absorption coefficient of distilled alcohol $\kappa_\lambda^A$ can be derived from the published absorbance $\cal{A}_\lambda$ or transmission $\cal{T}_\lambda$ data via
\begin{equation}
    ~~~~~~~~~~~~
    {\cal{A}}_\lambda \equiv \ln {\cal{T}}_\lambda = \rho_{\rm A}\, \kappa_\lambda^A \,L \ ,
\end{equation}
where $\rho_{\rm A}$ is the mass-density of the probe, $\kappa_\lambda^A$ is the absorption coefficient to be determined, and $L$ is the path length through the probe.
The mass-density of all alcohols in the laboratory are assumed to be that of ethanol: 0.789\,g/cm$^3$ (wikipedia).
The opacity in the gaseous phase is assumed to be identical to that in the liquid phase.

\begin{figure}
    \includegraphics[width=8cm]{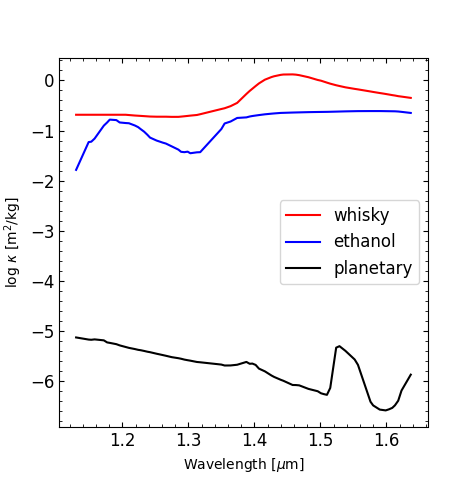}
    \caption{Mass-absorption coefficients for whisky (top), ethanol (middle) and a dry planetary atmosphere (bottom; from Swain et al. (2021)).}
    \label{fig:absorption}
\end{figure}

We extracted the mass-absorption opacity of pure ethanol from a figure found in the \cite{wiki:ethanol}.
The contributor used a fibre-feed Ocean Optics spectrometer (NIR-512) with a temperature-regulated InGaAs detector; the light from a halogen light bulb was shown through a beaker of liquid ethanol ($L$=2\,cm).
The result is shown in Fig.\,\ref{fig:absorption}: the opacity is very high compared with that of a dry planetary atmosphere (also shown in Fig.\,\ref{fig:absorption}); the opacity drops dramatically below 1.2\,$\mu$m but remains high for $\lambda > 1.4\,\mu$m; and there is a local minimum around 1.3\,$\mu$m.

For alcohol subjected to additional technical processing, we use the NIR spectra of \cite{6126980} covering a range of whiskys.
A Perkin Elmer spectrophotometer in the wavelength range 290-1880\,nm at 5\,nm resolution. 
The whisky probe was contained in a Hellma QS quartz cuvette ($L=1\,$mm).
The opacity is even higher than before (Fig.\,\ref{fig:absorption}) but this time the spectrum shows a prominent broad peak around 1.4\,$\mu$m instead.
We adopt the highlands probe over the islands probe; the differences are small in the NIR.

The relatively high opacities relative to that usually expected imply that searches for distilled alcohol are quite sensitive to small quantities, and the peaks/drops in the opacities between 1.2 and 1.5\,$\mu$m provide distinct spectral features to search for.

\section{Adding distilled alcohol to models of exoplanet transmission spectra}

The effort needed to produce a good opacity model for transmission spectra is considerable.
However, since we only desire to add a minor chemical constituent -- distilled alcohol -- we present here a genial method for perturbing a known model.

\cite{2017MNRAS.470.2972H} presented an analytic model for the effects of an isothermal planetary atmosphere on the transit lightcurves of exoplanets.
The effective transit cross-section of the atmosphere at a wavelength $\lambda$ is
\begin{equation}
    ~~~~~~~~~~~~
    {\tt A}_\lambda = \pi\, R_0^2 + 2\,\pi \,\int_{R_0}^\infty \left( 1 - \frac{I_\lambda(R)}{I_{\lambda,*}}\right)\, R \,dR = \pi\, R_\lambda^2 \ ,
    \label{eqn:area}
\end{equation}
where $R_0$ is a reference radius roughly corresponding to that of the opaque surface, $I_\lambda(R)/I_{\lambda,*}$ is the fractional reduction in transmitted stellar intensity -- for a simple two-stream model $e^{-\tau_\lambda}$ -- and $R$ is the cylindrical impact parameter centered on the planet.
For an isothermal atmosphere, this equation can be solved for the effective radius of the star at wavelength $\lambda$,
\begin{eqnarray}     \label{eqn:radius}
    ~~~~~~~~~~~~
    R_\lambda & \!\approx\! & R_0 + H \,\left( \gamma + \ln(\tau_0) + E_1(\tau_0) \right) \\ \nonumber
    \tau_0 & \!\approx\! & \frac{P_0 \,\kappa_\lambda}{g_0} \sqrt{\frac{2\,\pi\, R_0}{H}} \nonumber
\end{eqnarray}
where $R_0$, $P_0$, and $g_0$ are the atmospheric reference radius, pressure and surface gravity, $H$ is the effective atmospheric scaleheight, $\gamma$ is the Euler-Mascheroni constant (approximately 0.5772), $\kappa_\lambda$ is the effective mean mass-absorption coefficient, and $E_1(\cdot)$ is the first elliptical integral \citep{2017MNRAS.470.2972H}. 
There are substantial parameter degeneracies in interpreting transmission spectra, but here we are only interested in detecting minor trace components of the atmosphere.

Given values of $R_\lambda$ and $R_0$ one can invert Eqn.\,\ref{eqn:radius} to obtain the opacities used to model the transmission spectra from the published model.
Since the choice of $R_0$ is arbitrary -- other than the fact that the broadband optical depth $\tau_0$ of a line-of-sight at $R_0$ should be large -- we set $R_0$ equal to the radius at which $\tau_0 = 10$ for the purpose of re-construction the opacities, i.e. $R_0 = R_{\rm p}+H(\gamma+1)$.
However, we will leave it as a free parameter for the fits using modified opacities that follow.

\section{GJ 1132\,b}

GJ\,1132\,b is a super-Earth-type exoplanet in a tight 1.6-day orbit around a cool M4.5 dwarf \citep{2015Natur.527..204B}.
With a mass of 1.66\,M$_\oplus$ and a radius of 1.16\,R$_\oplus$ \citep{2018A&A...618A.142B}, the mean density is 6.3\,g\,cm$^{-3}$, making it a massive terrestrial object -- a ``super-Earth''.
The short orbital period suggests that the planet rotates synchronously with its orbit \citep{2017CeMDA.129..509B}.
A second and slightly more massive ``super-Earth'' is also present in the system -- GJ\,1132\,b -- in a larger 8.93-day orbit near the habitable zone \citep{2018A&A...618A.142B}.

The thermal equilibrium temperature of GJ\,1132\,b is about 529\,K.
With such a high mean equilibrium temperature, the planet should have lost its primordial gaseous atmosphere.
Nevertheless, \cite{2021AJ....161..213S} claimed a detection of an atmosphere using HST infrared grism transit spectroscopy; they suggest that there must be enough volcanic outgassing to maintain a substantial atmosphere despite continuing atmospheric mass-loss.
Although the side of the planet facing its host star is undoubtedly hot, the fact that the planet is synchronously rotating suggest a colder night-side temperature of about
\begin{equation}
    ~~~~~~
    T_{\rm night} \approx T_{\rm eq} \left(\frac{\kappa \,P}{g}\right)^{\frac{1}{4}}
    ~ = ~ 300\,{\rm K} ~ \left(\frac{\kappa}{10^{-5}\,{\rm m}^2\,{\rm kg}^{-1}}\right)^{\frac{1}{4}} \left(\frac{P}{1\,{\rm bar}}\right)^{\frac{1}{4}}
\end{equation}
\citep{2015ApJ...806..180W}, which makes the night-side roughly habitable. 
Note that \cite{2023ApJ...959L...9M} used the James Web Space Telescope to obtain two IR transmission spectroscopy transits of GJ\,1132\,b in the spectral range 2-5\,$\mu$m, one of which was consistent with an atmosphere with water, so a purely H$_2$ atmosphere may not be the case.
However, their inability to obtain a consistent spectral signal means that we will only consider the Hubble NIR data here.

\begin{figure}
    \includegraphics[width=9.3cm]{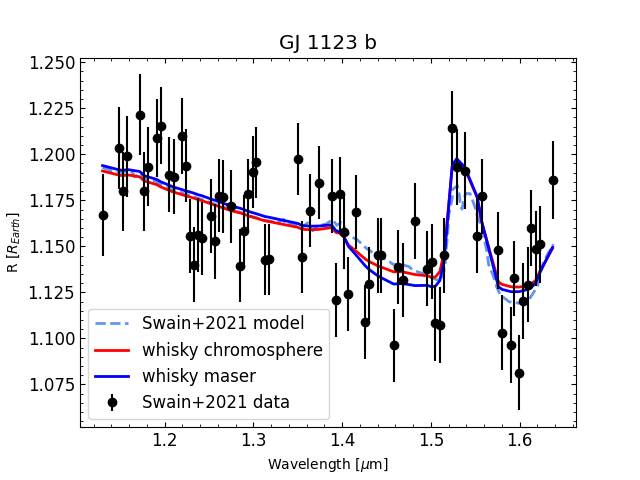} 
    \caption{Residual transit areas
    from Swain et al.~(2021); 
    their fit is shown as a solid blue line; the fit including alcohol is shown as a solid red or blue line.}
    \label{fig:data}
\end{figure}

The NIR transmission spectra and calibration data analysed by \cite{2021AJ....161..213S} were obtained during 5 visits of 4 orbits each and extracted later by the authors from the MAST archive.
A detailed account of the reduction, analysis, and fitting of the transit data can be found in their paper.
The residual radii $R_\lambda-R_0$ derived from the published $\Delta R^2/R_0^2$ are shown as data points in Figure\,\ref{fig:data}.
Visible is a slight decrease in size towards larger wavelengths and an emission line-like feature  at about 1540\,nm corresponding to HCN, as well as suggestions of CH$_4$ at about 1390\,nm and perhaps beyond 1600\,nm.
The increase in size below 1500\,nm can be explained as due to Rayleigh scattering.

The atmospheric model fitted by \cite{2021AJ....161..213S} consisted of a predominantly H$_2$ atmosphere plus small traces of aerosols like CH$_4$ and HCN -- the latter an indication of photochemistry.
They argue that the previously claimed detection of water by \cite{2017AJ....153..191S} using data obtained from the Earth's surface rather than in space is incorrect.
Given the presence of methane and molecular hydrogen, it is not unreasonable to suspect that ethanol (C$_2$H$_5$OH) could also survive.

Given the extreme irradiation of the daylight side of the planet, one can argue that a simple absorption model for the brightness of the planetary limb is too simple.
If the irradiation is strong even at the limb -- only the limbs are used during the transit to measure the effective planetary radii -- one might expect to see the formation of a chromospheric layer.
When highly reactive species like ethanol reach the illuminated limbs of the planetary atmosphere, they will be heated, first resulting in emission and eventually in ionization/combustion.
We call this type of chromosphere a {\it flamb\'e-ed} atmosphere.
A simple model for the limb-brightening due to the formation of a chromosphere is derived in the Appendix.

\section{Results}

\begin{table*}
    \caption{~~~~~~~~~~~ Results}
    \begin{tabular}{|l|ccccccc|}
        \hline
        Model & $R_0$ & $\Delta R_{\rm HCN}$  & $\log(f_{\rm ethanol})$ & $\log(f_{\rm whisky})$ & $B_{\rm A}/I_\star$ & reduced $\chi^2$ \\
              & $\left[ R_\oplus \right]$ & $\left[ R_\oplus \right]$ & & & & &  \\
        \hline
        Swain et al. (2021) & ($1.16$) & -- & -- & 
            -- & -- & ($1.26$) \\
        incl. ethanol 
            & $1.11 \pm 0.02$ & $< 0.10$ & $< -7.1$ & - & $\equiv 0$ & $1.39$ \\
        incl. whisky 
            & $1.11 \pm 0.06$ & $< 0.09$ & -- &  $< -6.8$ & $\equiv 0$ & $1.45$ \\
        incl. ethanol chromosphere 
            & $1.12 \pm 0.02$ & $< 0.01$ & $< -4.0$ & -- &  $\equiv 1$ & $1.39$ \\
        incl. whisky chromosphere
            & $1.121 \pm 0.006$ & $0.023 \pm 0.015$ & - &  $-6.3 \pm 0.7$ & $\equiv 1$ & $1.16$ \\
        incl. whisky maser
            & $1.124 \pm 0.005$ & $0.022 \pm 0.012$ & - &  $-7.0 \pm 0.2$ & $\equiv 3$ & $1.12$ \\        \hline
    \end{tabular}
    \label{tab:results}
\end{table*}

We fit our distilled alcohol model to the same NIR transmission spectra using Eqs.\,\ref{eqn:radius}. 
The main fitted parameters are $R_0$ and the logarithm of the relative alcohol abundance, $\log(f_{\rm A})$. 
In order to avoid having to deal with the complex chemistry responsible for the production of the observed HCN emission (we will ignore the much weaker CH$_4$ lines), we also fit a generic HCN correction to the final radii obtained with our model using a simple Gaussian-Squared line with known position and width but with unknown strength $\Delta R_{HCN}$ that matches the line present in the model by \cite{2017AJ....153..191S}.
Those interested in HCN are then welcome to convert the radial correction into modified HCN abundances.

After extracting the basis opacity by inverting Eqn.\,\ref{eqn:radius}, we re-fit the radius measurements using the Metropolis Monte Carlo sampler {\tt emcee} \citep{2019zndo...3543502F}, yielding robust estimates and uncertainties or upper limits for the parameters.
The initial results for a non-chromospheric atmosphere with either an ethanol or whisky opacity contribution are shown in Table \ref{tab:results}: no alcohol of any kind could be detected, even at a very low level.

Fortunately, the residual area data (Fig.\,\ref{fig:data} show a depression exactly in the region where the whisky opacity shows a distinct peak.
A higher opacity causing a stronger absorption would normally increase the effective radius rather than decreasing it, but this is no longer true if the medium is optically thin and in emission: the more emission present in the form of a limb-brightening, the more the atmosphere looks like the un-absorbed host star, decreasing the effective radius of the planet.
Thus, we also fit a chromospheric ``flamb\'e-ed atmosphere'' model (derived in the Appendix) using a simple least-squared optimization to fit the data.
We assume that the irradiated atmosphere at the edge of the planet is fully chromospheric (the parameter $f_c \equiv r_c/R_0$ introduced in the Appendix is assumed to be unity) and in thermal equilibrium with the stellar radiation field, so that the chromospheric source function of the distilled alcohol, $B_{\rm A}$, is equal to the stellar intensity (see the Appendix for details).
The results for the ``flamb\'e-ed'' atmosphere are also shown in Table\,\ref{tab:results}: no ethanol could be detected but a clear whisky contamination was detected at a level of 0.53 ppm.
This alcohol is seen mainly as a depression of the fitted radii around the whisky emission feature around $1.4\,\mu$m due to chromospheric emission.

The fitted relative density is quite high: the implied total mass of whisky in the un-illuminated half of the planetary surface is roughly
\begin{eqnarray}
    M_{\rm A} & \!\approx\! &
        \frac{1}{2} \, f_{\rm A} \, 4\,\pi \int_{R_{\rm p}}^\infty 
        \frac{\mu \, m_H \, P_0}{k\,T_0} e^{-\frac{r-R_{\rm p}}{H}} \,r^2 \,dr \\ \nonumber
        & \!\approx\! & 1.5\times10^{13}\,{\rm kg} \,\left(\frac{P_0}{10\,{\rm bar}}\right) \left(\frac{T_0}{480\,{\rm K}}\right)^{-1} 
        \ .
\end{eqnarray}

It is possible to achieve yet a better fit by increasing the effects of limb-brightening.
For instance, if $B_c/I_\star$ was as large as 3, one gets the fit labeled ``whisky maser'' in Fig.\,\ref{fig:data} (the fitted parameters are also listed in Table\,\ref{tab:results}). 
Such a fit simultaneously reduces the needed density of whisky to 0.11 ppm or by a factor of nearly 5.
Whether stimulated emission is possible is a question that thus needs to be addressed.

Before trying to identify the origin of the alcohol, one must consider that he atmosphere along the edge of the illuminated co-rotating planet must be dominated by extreme atmospheric flows driven by the one-sided heating of the planet \citep{2015ApJ...806..180W, 2023MNRAS.526..263B}.
The resulting atmosphere consists of a hot component that cools as it moves into the night-side and a corresponding cooler component that returns the atmosphere back to the side facing the host star.
The observed alcohol in GJ\,1132\,b can't survive long in the hot flow, given the spontaneous combustion temperature of about 365\,K at 1\,atm, so the alcohol must be produced on the cooler side and advected to the hotter side, where it combusts. 

The amount of whisky production can be estimated by assuming that there is a stationary flow of whisky-laden atmosphere towards the daylight side of the planet (accompanied by a returning flow where the whisky has been burned).
With a typical flow velocity of $\sim 10\,$m/s \citep{2015ApJ...806..180W},
\begin{equation}
    \dot{M}_{\rm A} \approx \frac{2\, M_{\rm A}\, u_{\rm atm}}{\pi\, R_{\rm p}} \approx 4\times10^{14}\,{\rm kg}\,{\rm yr}^{-1} \ .
\end{equation}
To place this number in perspective, it is $\sim$1400 times more than the global annual production of alcoholic drinks on Earth in 2023 \citep{statista}.

\section{The origin of the alcohol}

The detection of considerable amounts of whisky in the atmosphere of GJ\,1132\,b was unexpected, since it is not normally found in the atmospheres of planets.
We consider several possible origins.

\subsection{Accretion}

Alcohol is present in very small quantities in interstellar space \citep{2022cosp...44.2739M} and in comets \citep{2015SciA....1E0863B}.
Although it is conceivable that simple alcohols could be accreted onto GJ\,1132\,b, it seems highly unlikely that whisky could be produced in interplanetary space or as a result of star formation; an astrophysical origin of the observed whisky can be ruled out.

\subsection{Non-equilibrium atmospheric chemistry}

The atmospheric model calculated by \cite{2021AJ....161..213S} was an equilibrium model that apparently did not predict any significant amounts of alcohol of any kind.
In any case, the chemistry is bound to be complex and non-equilibrium: would it be possible to produce a whisky-like alcohol using this mechanism?

The chemical process responsible for the basis alcohol must either be hydrogenation -- combining water with an alkene (unsaturated hydrocarbons of the form C$_n$H$_{2n}$ or C$_n$H$_{2n-2}$) -- or hydration, which adds hydrogen to a ketone or an aldehyde, the latter consisting of C=O carbonyl groups.
However, the relative lack of water suggests there is also a lack of oxygen, so it seems that non-biological processes are unlikely to be taking place to produce the basis-alcohol, much less to refine it to a primitive whisky. 

\subsection{Primitive biological activity}

It is possible that there are simple biological organisms like yeast cells on the surface or even in the atmosphere of GJ\,1132\,b capable of producing alcohol and perhaps processing it so efficiently that it resembles distilled whisky.

Atmospheric yeast is unlikely to be present in the hot atmospheric flow, since terrestrial yeast cannot withstand such temperatures.
If atmospheric yeast is grown on the cooler side and subsequently destroyed as it is transported to the hot side, then there must be a process by which the yeast is replenished.
Without the presence of substantial amounts of atmospheric sugar, it seems unlikely that enough yeast would be produced.
Thus, the yeast can only be on the surface.

The mass of yeast required can be easily estimated; the mass of needed to produce a bottle of beer or wine is -- astronomically speaking -- about the same as the mass of the resulting alcohol, so terrestrial yeast can produce alcohol at a rate of about 1\,liter/kg/week.
Given a total alcohol production on GJ\k1132\,b of $4\times10^{14}$\,kg/yr, we require
$\sim 8\times10^{12}\,$kg of yeast, equivalent to a uniform layer of only $140\,\mu m$ over the entire surface assuming the same density as sour-dough bread \citep{aqua-calc}.

Although this scenario would appear to be possible, there remains the question of how such yeast would be fed and how the additional substances that change ethanol to whisky would be added.

\subsection{Extraterrestrial civilisation}

If we can use our own civilisation as a model, there is no doubt that alcohol can be produced -- even in large quantities -- by a reasonably technical civilisation.
Indeed, the production of alcohol may be the critical step needed to transform a hunter-gatherer culture to a more settled form \citep{dietrich_heun_notroff_schmidt_zarnkow_2012}.
As pointed out earlier, the production of {\it atmospheric} whisky corresponds to $\sim 1400$ times that of the current global production of alcoholic beverages, but once a reasonable industrial capacity has been reached -- say, that of the industrial revolution of the 19th century -- it is only a question of how much effort is put into the production and how much of the alcohol escapes into the atmosphere.

\subsection{The most likely source}

We conclude that the only reasonable way to produce so much alcohol is via the technical processes of a reasonably advanced extraterrestrial civilisation.

A civilisation with origins on GJ\,1132\,b would have gone through a difficult evolutionary path:  the original life may have formed while the planet still had its primordial atmosphere and have adapted later to the more terrestrial form seen now or it may have only appeared after the removal of the primordial gas.
The alternative is that life and the civilisation actually originated on GJ\,1132\,c -- a much more habitable planet -- and the whisky distilling was transferred much later to the more inhospitable GJ\,1132\,b.
The latter would require that the civilisation at least reached a technological level capable of interplanetary travel between the two planets -- an aspect of additional relevance, as we will see.

Why would a reasonably technically advanced civilisation produce so much atmospheric alcohol?
There are several possible explanations:
\begin{itemize}
    \item The atmospheric whisky could have been released as an industrial accident of planetary proportions.  This seems unlikely for several reasons, including the fact that the global atmospheric circulations driven by the day-side heating of the planet would quickly transfer the alcohol to where it would be naturally burned and removed.  We would then have accidentally observed this singular event exactly when it occurred.
    \item A more likely explanation is that the atmospheric whisky is caused by routine industrial pollution due to leaks during production, storage, and/or transfer.  Assuming a reasonable leak-fraction of 1\%, the total production of alcohol would rise to about $4\times10^{16}$\,kg/yr.
\end{itemize}
Either way, we must seek the rationale for the extremely high production rates -- rates that far exceed what one would normally expect.
\begin{itemize}
\item One option is that the alcohol is needed for the extraterrestrial economy, for example as rocket fuel.
This idea would support the expectation that the civilisation started on the more habitable planet GJ\,1132\,c and runs a planetary rocket-fuel industry on GJ\,1132\,b to support that civilisation.
However, wouldn't it be far easier to extract the atmospheric hydrogen suggested to be present by \cite{2021AJ....161..213S} rather than attempt to distill alcohol?  How could it be that whisky is a better rocket fuel than ethanol?
\item The more reasonable option is that the whisky is needed for personal consumption and that industrial distillation on a massive planetary scale is necessary to fulfill the demand.  If so, the consumption must be very high:
for a population of 10 billion extraterrestrials, the annual use would be $(4\times10^{14}-4\times10^{16}$\,litres)/$10^{10}$ or 40-400 thousand liters/alien/year.
It is perhaps no wonder that the amount of industrial whisky pollution is so great.
\end{itemize}


\section{Can we characterize the extrasolar planet's climate?}

The different spectroscopic signature of whisky relative to ethanol is due to the ``peatiness'' of the distilled mash used.
Peat is produced in a particular climate characterized by a low mean temperature and considerable rainfall onto unforested landscapes \citep{keddy}.
Thus, a high ``peatiness''-index would suggest a cool, wet climate whereas a low ``peatiness''-index would be more characteristic of a warmer, drier climate.
It may be possible that the choice of whisky over purer forms of alcohol like vodka is a reflection of the planet's climate: relatively cool and damp.

Given the possible complexity of alcohol-contaminated planetary atmospheres, it is possible that atmospheric retrieval models may have a hard time recovering the molecular features on worlds where the peatier ``island'' varieties are favoured. Such planets are more likely to have upper atmospheric haze layers of complex hydrocarbons generated by peat burning, producing a comparatively featureless transit spectrum, cf. the JWST/MIRI results of \cite{2023ApJ...951...96G} for the hot Neptune GJ\,1214\,b which suggest a high molecular weight and possibly non-hydrocarbon, non-nitrogen haze precursors.

Conversely, the civilisation on GJ\,1132\,b may have discovered a way to age distilled alcohol under totally different climatic conditions using the abundant volcanic out-gassing responsible for the observed atmospheric composition rather than with peat and the usual oaken casts.  
This solution would go a long way towards explaining their ability to produce so much alcohol.
If so, a totally new whisky production method may be possible on Earth at places like Hawai'i and Iceland -- further research in this direction is necessary \citep{2023arXiv230400319H}.

\section{Implications for other extraterrestrial civilisations}

We now know that most stars have planets, that many of these planets are within the habitable zones of their host stars, and that there is a good chance that many of the latter harbour some form of life.
Although only a tiny fraction of these are likely to have developed technical civilisations, the shear number of planets within our galaxy practically guarantees that there should be a large number of detectable technically advanced civilisations.
Since those are unlikely to be able to hide their presence, one is inclined to ask
\begin{quotation}
\noindent
``... but-a where-a is-a everybody?''
\end{quotation}
During the summer of 1950, the famous physicist Enrico Fermi was walking with his colleagues Emil Konopinski, Edward Teller, and Herbert York, discussing recent UFO reports and the associated inference that faster-than-light travel would have to be possible.
If a large number of extraterrestrial civilisations should be capable of contacting the Earth or at least showing detectable radio emissions not seen by dedicated searches like SETI \citep{2001ARA&A..39..511T}, then why haven't we seen something? -- this is the so-called ``Fermi-Paradox'' \citep{2014arXiv1403.8146K}.

Previous discussions have assumed that a large number of advanced technical civilisations should be detectable and the lack of detection suggests that these are not present in our Galaxy, but there is a third possibility -- there may be civilisations that have been stunted in their technological development.
Given the large amount of whisky in the atmosphere of GJ\,1132\,b, there can be no doubt that the technical ability of this civilisation must be hindered somewhat by the high levels of consumption -- they are simply not up to the intellectual challenges associated with interstellar travel and maybe even radio technology.
It is also questionable if such a civilisation is capable of interplanetary travel of such magnitude as to permit the export of whisky from GJ\,1132\,b to c; the suggestion is that this extraterrestrial civilisation has its origins on the inner planet and only need to reach the industrial capacity of 19th or early 20th century Europe/USA.
As such, it would be difficult to detect in routine SETI observations -- we suggest a concerted global effort to set limits on the total technical radio signals emanating from GJ\,1132\,b.

Further detections of high amounts of distilled alcohol in other exoplanet atmospheres would provide a generic solution to the Fermi-paradox.
Until then, it remains to be seen how many other civilisations are trapped by this effect.
We have submitted a proposal to look for the spectroscopic signs of tetrahydrocannabinol in far IR transmission spectra \citep{joint} as a promising additional probe of this phenomenon.

\section{Conclusions}

We have detected the presence of large amounts of whisky in the atmosphere of the extrasolar planet GJ\,1132\,b using previously obtained transmission spectroscopy in the NIR.
The alcohol is unlikely to be of natural origin -- accretion, non-biological or naturally occurring yeast -- so that the only remaining explanation is that there is a technically advanced civilisation on the planet.
The simplest explanation for the high levels of atmospheric whisky is that it is due to industrial pollution, probably from the production of alcohol meant for personal consumption.
The quantities of alcohol are so great that any such civilisation would probably be incapable of the extraordinary challenges associated with extensive radio emissions, much less interstellar travel, potentially explaining the Fermi Paradox.

\section*{Acknowledgements}

All but the last paragraph of the introduction was written by {\tt ChatGPT} using the public web-interface provided by {\tt openai.com}.

This research was supported by the International Science Partnership grant {\it Buaidh Astrofiosaig Uisge-Beatha Braiche} (\#2023-astro-45\%) of the {\it Comhairle Maoineachaidh na h-Alba}.



We made use of the tools provided by the public domain {\tt numpy},  {\tt astropy}, {\tt emcee} and {\tt matplotlib} python libraries, as well as commercial but here unnameable sources of whisky.  

\section*{Data Availability}
 
This work makes use of public data.
The transmission spectroscopic data was provided by NASA's {\it Exoplanet Archive}.
The distilled alcohol absorbances and the theoretical radii from \cite{2017AJ....153..191S} were extracted from the published figures using screen shots and the ``Figure Calibration'' plugin to ({\tt Astro}){\tt ImageJ} available from the authors.
The resulting tables and the python script used to fit the data are available from the authors upon request.



\bibliographystyle{mnras}
\bibliography{april1}



\newpage
\appendix

\section{The effects of a planetary chromosphere on transit radii}

We adopt the basic model of \cite{2017MNRAS.470.2972H} but make a few corrections to their notation (e.g. Eqn.\,\ref{eqn:area} is in cylindrical, not radial coordinates) and derivation.
All opacities, optical depths, and the derived planetary radii are, of course, wavelength-dependent, but we will drop wavelength from the notation for clarity.

In an isothermal model, the atmospheric mass-density is assumed to be
\begin{equation}
    ~~~~~~~~~~~~
    \rho(r) = \rho(R)\, e^{-\frac{r-R}{H}} = \rho(R)\, e^{-\frac{\Delta r}{H}}  \ ,  
\end{equation}
where $r=R+\Delta r$.
The optical depth along the observer's line-of-sight through the atmosphere along a perpendicular coordinate $x$ is
\begin{equation}
    ~~~~~~~~~~~~
    \tau(R) = \int_{-\infty}^{+\infty} \rho(R(x))\, \kappa \, dx \ ,
\end{equation}
where the mass-absorption coefficient $\kappa$ is assumed to be independent of $r$.
To express the radius $R(x)$ in terms of $R$ and the line-of-sight variable $x$, we make the usual assumption that $x^2 \equiv r^2-R^2 = (R+\Delta r)^2-R^2 \approx R^2+2\,R\,\Delta r$ so that $\Delta r = r-R \approx x^2/2\,R$, yielding
\begin{eqnarray}     \label{eqn:tau}
    \tau(R) & \!\approx\! & 2\, \kappa\, \rho(R) \,\int_0^{+\infty} e^{-\frac{x^2}{2\,R\,H}} \,dx \\ \nonumber 
    & \!\approx\! & \sqrt{2\,\pi\, H\, R}\, \kappa \,\rho(R) \\ \nonumber
    & \!\approx\! & \tau_0 ~ \sqrt{\frac{R}{R_0}} ~ e^{-\frac{R-R_0}{H}} \\ \nonumber
    \tau_0 & \!\equiv\! &
        \sqrt{2\,\pi \,H \,R_0} \,\kappa \,\rho(R_0) \ .
\end{eqnarray}

For a normal atmosphere, the effective planetary area is (Eqn.\,\ref{eqn:area})
\begin{equation}
    ~~~~~~~~~~~~
    \label{eqn:area2}
    {\tt A} = \pi\, R_0^2 + 2\,\pi \,\int_{R_0}^\infty \left( 1 - e^{-\tau(R)} \right)\, R \,dR \ .
\end{equation}
This integral can be solved approximately by noting that
\begin{equation}
    ~~~~~~~~~~~~
    \frac{d\tau}{dR} = \tau \,\left( \frac{1}{2} \frac{1}{R} - \frac{1}{H} \right) \approx -\frac{\tau}{H} \ .
\end{equation}
By re-formulating the integral from using $R$ to $\tau$, one obtains Eqn.\,\ref{eqn:radius}, the classic result discussed in detail by \cite{2017MNRAS.470.2972H}.

When one adds a planetary chromosphere with different optical properties, such a simple analytic solution is no longer possible.
We assume that an additional opacity component ``A'' is present throughout the atmosphere with a relative density $f_{\rm A} \equiv \rho_{\rm A}/\rho$ but which additionally results in chromospheric emission for all radii $r > r_c$ with an assumed constant source function $B_{\rm A}$.
Let
\begin{equation}
    \epsilon_{\rm A} \equiv \frac{\kappa_{\rm A} \,f_{\rm A}}{\kappa_{atm}+\kappa_{\rm A}\, f_{\rm A}} \equiv \frac{\kappa_{\rm A}\, f_{\rm A}}{\kappa}
\end{equation} 
be the relative contribution of opacity due to A so that $\epsilon_{\rm A} \tau$ is the contribution to the optical depth $\tau$ solely due to A.

Lines-of-sight with $R > r_c$ are wholly within the chromosphere, so the emergent intensity $I(R)$ in the two-stream approximation is
\begin{equation}
    ~~~~~~~~~~~~
    I(R \ge r_c) = I_\star\, e^{-\tau(R)} + B_{\rm A} \,\left( 1 - e^{-\epsilon_{\rm A} \tau(R)} \right) \ ,
\end{equation}
where $\tau(R)$ is calculated normally using Eqn.\,\ref{eqn:tau}.

For lines-of-sight with $R \le r_c$, one must consider the effects of a line-of-sight that first transits part of the chromosphere, then enters the purely absorbing atmosphere when
\begin{equation}
    ~~~~~~~~~~~~
    x = x_c \equiv \sqrt{r_c^2-R^2}
\end{equation}
and re-enters the chromosphere again at $x \! = \! -x_c$.
Let $\tau_a(R)$ be the total optical depth within the non-chromospheric atmosphere and $\tau_c(R)$ be the total chromospheric optical depth.  
Thus,
\begin{eqnarray}
    \tau_{a}(R < r_c) & \!=\! & 
        2 \,\int_{0}^{x_c(R)} \kappa \,\rho(R_0)\, e^{-\frac{r-R_0}{H}} \,dx \\ \nonumber
    & \!=\! & 
        \sqrt{2\,\pi\, H\, R} \, \kappa\, \rho(R_0) \, {\rm erf}\left(\sqrt{\frac{r_c^2-R^2}{2\,R\,H}}\right) \\ \nonumber
    & \!=\! & 
        \tau_0 ~ \sqrt{\frac{R}{R_0}} ~ {\rm erf}\left(\sqrt{\frac{r_c^2-R^2}{2\,R\,H}}\right) \ .
\end{eqnarray}
Similarly,
\begin{eqnarray}
     \tau_c(R < r_c) & \!=\! & 
        2\, \int_{x_c(R)}^{\infty} \kappa\, \rho(R_0) e^{-\frac{r-R_0}{H}}\, dx \\ \nonumber
        & \!=\! & 
        \tau_0 ~ \sqrt{\frac{R}{R_0}} ~ {\rm erfc}\left(\sqrt{\frac{r_c^2-R^2}{2\,R\,H}}\right) \ .\nonumber
\end{eqnarray}
Now, the emergent intensity in the two-stream approximation is
\begin{eqnarray}
    I(R < r_c) & \!=\! &
            \left[ I_\star\, e^{-\frac{1}{2}\tau_c} 
        + B_{\rm A} \,\left(1-e^{-\frac{1}{2} \epsilon_{\rm A} \tau_c}\right) \right]\, e^{-\tau_{a}}
            \\ \nonumber
    & & ~~~~~~~~~~~~~~~~~~~~
        + B_{\rm A}\, \left(1+e^{-\frac{1}{2}\epsilon_{\rm A} \,\tau_c}\right) \\ \nonumber
    & \!=\! &
        I_\star\, e^{-\tau} 
        + B_{\rm A} \, \xi(\tau_a,\tau_c,\epsilon_{\rm A}) \\ \nonumber
    \xi(\tau_a,\tau_c,\epsilon_{\rm A}) & \!\equiv\! &
        \left(1-e^{-\frac{1}{2} \epsilon_{\rm A}\, \tau_c}\right) e^{-\tau_a-\frac{1}{2} \tau_c} \\ \nonumber
    & & ~~~~~~~~~~
        + \left(1+e^{-\epsilon_{\rm A}\, \tau_a}\right) e^{-\frac{1}{2}\tau_c} 
        + \left(1+e^{-\frac{1}{2}\epsilon_{\rm A}\, \tau_c}\right) \ .
        \nonumber
\end{eqnarray}
Finally, we obtain the equation for the effective planetary area,
\begin{eqnarray}
    {\tt A} & \!\approx\! & \pi\, R_0^2 \\ \nonumber
        & & + ~
        2\,\pi \int_{r_c}^\infty 
        \left(1-\frac{B_{\rm A}}{I_\star}\right)
        \left(1 - e^{-\epsilon_{\rm A}\, \tau_c(R)}\right)
        \,R \,dR \\ \nonumber
    & & + ~
        2\,\pi\, \int_{R_0}^{r_c} \left(
        1 - e^{-\tau(R)} 
         -  
        \frac{B_{\rm A}}{I_\star}\, \xi(\tau_a,\tau_c,\epsilon_{\rm A})
        \right)\, R \,dR\ . \nonumber
    \label{eqn:final_area}
\end{eqnarray}
The corrected effective area thus depends upon two additional parameters, $r_c$ and $B_{\rm A}/I_\star$.
Since $r_c$ must necessarily scale in some way with $R_0$, we chose to parameterize this radius as some fraction $f_c \equiv r_c/R_0 > 1$ of the latter.

\bsp	
\label{lastpage}
\end{document}